\documentstyle[preprint,aps]{revtex}
\begin{document}
\draft
%
\title{
Flux String in Quantum Billiards 
with Two Particles
}
\author{
{Taksu Cheon}
}
\address{
Department of Physics, Hosei University, Fujimi, 
Chiyoda-ku, Tokyo 102, Japan
}
\author{
{T. Shigehara}
}
\address{
Computer Centre, University of Tokyo, Yayoi, 
Bunkyo-ku, Tokyo 113, Japan 
}
%
\date{March 14, 1996}
\maketitle
\begin{abstract}
We examine the quantum motion of two particles
interacting through a contact force which are
confined in a rectangular domain in two and three dimensions.
When there is a difference in the mass scale
of two particles, adiabatic separation of
the fast and slow variables can be performed.  
Appearance of the Berry phase and magnetic flux 
is pointed out. 
The system is reduced to a one-particle Aharonov-Bohm 
billiard in two-dimensional case.
In three dimension, the problem effectively becomes the motion of a
particle in the presence of closed flux string in a box billiard. 
\end{abstract}
%
%
\narrowtext
%

The quantum billiard problem has been instrumental in
revealing non-trivial features of seemingly simple
systems with minimal degrees of freedom.
As a consequence, unlike real-world billiards,
quantum physicists' version of the billiard problem
has only one particle moving within  
a walled boundary.
The dynamics is controlled by the shape of
the boundary, and by optional obstacles placed
inside the billiard \cite{MK79,BG83,BW84,AS88,CC89}.
The formulation of zero-size obstacle \cite{AG88,SE90} offers a
possibility to go beyond this limitation,
since it can be thought of as the second particle
with infinitely heavy mass.

In this Letter, we consider the quantum mechanics of two
and three dimensional systems consisting of two particles, 
one heavier than the other, interacting through a contact 
force, and confined in a walled rectangular boundary.
Adopting the adiabatic separation of fast motion of the
light particle and slow motion of the heavy one, we
can formulate the problem in two steps.  Namely, we first
solve the motion of the light particle with the location of the
heavy particle fixed, then calculate the effective Hamiltonian
for the heavy particle using the eigenstates of the light 
particle. The first step yields a one-particle 
billiard problem in the presence of a pointlike scatterer,
which has been the focus of several recent studies
\cite{SE90,SH94,CS96,SC96}.
The identification of the Berry phase \cite{BE84}
in this one-particle billiard \cite{CS96} then has an 
interesting ramification in the second
step: The effective Hamiltonian for the heavy particle acquires
a non-trivial vector potential which results in a field formally
identical to the magnetism.
We show that, in two dimension, this 
``magnetic'' field appears as infinitely thin flux lines penetrating 
the billiard.  Through a numerical example, we demonstrate 
that the flux lines have discernible effects both on the spectra 
and wave functions of the system.
We also show that, in three dimension, the flux lines form
closed strings which reside inside the billiard.  Several 
examples of the flux string are displayed for a box billiard.
  
%
Let us suppose that two particles $m$ and $M$ each located at 
$\vec r$ and $\vec R$
interacting through a delta force are confined in
a $d$-dimensional rectangular domain by a hard wall $V_{wall}$.
We assume that $M$ is substantially larger than $m$ so that
the motion of $M$ can be treated adiabatically.   
The Hamiltonian describing the system is given by 
%
%
\begin{eqnarray}
\label{1}
H = {{\vec P^2}\over{2M}}+h(\vec R)+V_{wall}(\vec R)
\end{eqnarray}
where $h(\vec R)$ is a sub-Hamiltonian for the light 
particle which is given by
\begin{eqnarray}
\label{2}
h(\vec R) = {{\vec p^2}\over{2m}}+v\delta ^d(\vec R-\vec r) 
  +V_{wall}(\vec r).
\end{eqnarray}
For a fixed value of $\vec R$, this sub-system is known 
as the billiard with a pointlike scatterer 
\cite{AG88,SE90,SH94}.  As it stands, the resolvent of this
sub-Hamiltonian is not well-defined for spatial dimension $d \geq 2$. 
However, in case of $d \leq 3$, the system can be
made meaningful by redefining a coupling constant $v$ 
in terms of renormalized coupling constant $\bar v$.
Assuming that the $h(\vec R)$ is handled with
such proper renormalization procedure \cite{SH94},
we write the eigenvalue equation of $h(\vec R)$ as 
\begin{eqnarray}
\label{3}
h(\vec R)\left|{\alpha(\vec R)}\right\rangle 
  = \omega_\alpha(\vec R)\left|{\alpha(\vec R)}\right\rangle.
\end{eqnarray}
There are two types of solutions to this equation \cite{CS96}.
Let us denote the eigenvalue and eigenstate with ${\bar v}=0$ 
(empty billiard) as $\varepsilon_n$ and $\left| n \right\rangle$. 
The first, trivial type of solution exists when the delta potential
is located on the node-line of unperturbed state;
\begin{eqnarray}
\label{4}
\omega_\alpha = \varepsilon_n  \ \ \ \   {\rm if} \ \ \ \
f_n(\vec R) = 0.
\end{eqnarray}
where $f_n(\vec r) = 0$ represents the node-line of the wavefunction
$\left\langle {\vec r} \right|\left. n \right\rangle$.  
The second, more generic type of solution of eq. (\ref{3}) 
is obtained from the equation
\begin{eqnarray}
\label{5}
\overline G(\vec R;\omega )-{1 \over {\bar v}}=0
\end{eqnarray}
where 
\begin{eqnarray}
\label{6}
\overline G(\vec r;\omega )=\sum\limits_n {
  \left\langle {\vec r} \right|\left. n \right\rangle \!\!
  \left\langle n \right|\left. {\vec r} \right\rangle 
  \left[ {{1 \over {\omega -\varepsilon _n}}
    +{{\varepsilon _n} \over {\varepsilon _n^2+1}}} 
  \right] }.
\end{eqnarray}
is the inverse of transition matrix of the light particle $m$. 
If we assume that $m$ stays at the adiabatic state 
$\left|{\alpha(\vec R)}\right\rangle$ during the motion of $M$,
we obtain an effective Hamiltonian governing the dynamics
of $M$ as 
\begin{eqnarray}
\label{7}
H^{e\!f\!f}_\alpha 
  &\equiv& \left\langle{\alpha(\vec R)}\right|H
           \left|{\alpha(\vec R)}\right\rangle \\ \nonumber 
  &=& {1\over{2M}}\left({\vec P-\vec A_\alpha(\vec R)}\right)^2
     +U_\alpha(\vec R)+V_{wall}(\vec R)
\end{eqnarray}
where the scalar potential $U_\alpha$ is 
given by
\begin{eqnarray}
\label{8}
U_\alpha(\vec R) = \omega_\alpha(\vec R),
\end{eqnarray}
and the gauge potential 
(Mead-Berry connection) $\vec A_\alpha$ is given
by \cite{BE84,MT79}
\begin{eqnarray}
\label{9}
\vec A_\alpha(\vec R)=i\left\langle{\alpha(\vec R)}\right|
               \vec \nabla_R
            \left|{\alpha(\vec R)}\right\rangle.
\end{eqnarray}
It is important that we choose
normalized eigenstate $\left|{\alpha(\vec R)}\right\rangle$ 
as a  single-valued function in the parameter space of $\vec R$. 
In some cases, $\vec A_\alpha$ can be made to be zero everywhere by
redefining the phase of the 
$\left|{\alpha(\vec R)}\right\rangle$, but in general,
this is not possible.
When $\vec R$ makes an adiabatic cyclic motion, it acquires a 
geometric phase given by the circular line integral of
$\vec A_\alpha$.
The appearance of the vector potential is a consequence of 
the fact that this phase is non-zero.
It is worth stressing that in the system considered here,
the parameter space of 
slow variable is nothing but the coordinate space of
heavy particle $\vec R$.  Thus the vector potential 
$\vec A_\alpha$ resides in the coordinate space $\vec R$.
The eigenvalue equation
\begin{eqnarray}
\label{10}
H_\alpha^{e\!f\!f}\left|{\Phi_{\alpha p}}\right\rangle 
  = E_{\alpha p} \left|{\Phi_{\alpha p}}\right\rangle
\end{eqnarray}
gives the energy states of the whole system $E_{\alpha p}$
along with the eigenstate of the heavy particle 
$\left|{\Phi_{\alpha p}}\right\rangle$.  They are indexed by
two integers $\alpha$ and $p$ each specifying the state for 
light and heavy particles.

We first concentrate on the case of two dimension.
Since the Hamiltonian is real symmetric,
the geometric phase is $\pi$ when the path of the cyclic 
motion of $\vec R$ surrounds
the diabolical point $\vec R^*$ \cite{BE84,HL63},
which is the point of degeneracy of the sub-Hamiltonian $h(\vec R)$; 
namely $\omega_\alpha(\vec R^*) = \omega_{\alpha \pm 1}(\vec R^*)$.
Otherwise, the phase is zero.
Since the
quantity $\overline G(\vec r; \omega)$ is a monotonous function of 
$\omega$ except at it's poles \cite{SH94}, the degeneracy can occur only
between the solutions of eqs. (\ref{4}) and ({5}).
Therefore, location of $\vec R^*$ is determined by
\begin{eqnarray}
\label{11}
f_n(\vec R^*) = 0
\ \ \ \   {\rm and} \ \ \ \
g_n(\vec R^*) = 0
\end{eqnarray}
where $g_n(R)$ is defined by
\begin{eqnarray}
\label{12}
g_n(\vec R)\equiv \overline G(\vec R;\varepsilon _n)
                 -{1 \over {\bar v}}.
\end{eqnarray}
In general, this equation has more than one solution for each 
degeneracy $\varepsilon _n = \omega_\alpha$. We write them as 
$\vec R_{\alpha j}^*$ $(j = 1, 2, \cdots)$.
Because of the sign reversions around $\vec R^*_{\alpha j}$, 
single-valued representation of the eigenfunction 
$\left|{\alpha(\vec R)}\right\rangle$ 
necessarily becomes complex even for the real symmetric Hamiltonian. 
There is a freedom to choose the gauge. 
We take the ``maximally symmetric'' choice around 
each $\vec R^*_{\alpha j}$;
\begin{eqnarray}
\label{13}
\left|{\alpha(\vec R)}\right\rangle = 
\exp{( \frac{i}{2} \sum\limits_j \varphi_{\alpha j})}
N_\alpha ({\vec R})
\sum\limits_n {
\left| n \right\rangle 
\frac{ \left\langle n | {\vec R} \right\rangle }
{\omega_\alpha({\vec R}) -\varepsilon _n}}
\end{eqnarray}
where $\varphi_{\alpha j}$ is an angle variable of $\vec R$ 
in the polar coordinate with its origin at 
$\vec R_{\alpha j}^*$, 
and $N_\alpha$ is a real normalization factor.  From 
eqs. (\ref{9}) and (\ref{13}), we obtain  
\begin{eqnarray}
\label{14}
\vec A_\alpha(\vec R)=\sum\limits_j {{-\vec e_{\varphi_{\alpha j}}} 
  \over {2\left|{\vec R-\vec R_{\alpha j}^*} \right|}}
\end{eqnarray}
where $\vec e_{\varphi_{\alpha j}}$ is a unit vector of the 
direction $\varphi_{\alpha j}$. We choose its direction such that 
$(\vec R-\vec R_{\alpha j}^*) \times \vec e_{\varphi_{\alpha j}}$ 
points to the positive $z$-axis in usual right-hand convention. 
The gauge curvature (or the ``magnetic'' field) $\vec B_\alpha$ is
given by 
\begin{eqnarray}
\label{15}
\vec B_\alpha(\vec R)=\vec \nabla \times \vec A_\alpha(\vec R)
  =\sum\limits_j  
   (-\vec n_z) \pi \delta ^2(\vec R-\vec R_{\alpha j}^*)
\end{eqnarray} 
where $\vec n_z$ is the unit vector of the direction of $z$-axis.
The problem is now turned into the motion of a particle
$M$ in the presence of infinitely thin
lines of magnetic flux piercing through the
billiard. This type of system is known
as Aharonov-Bohm billiard \cite{BR86,DJ95}.
Each flux line has the strength $\pi$, or 1/2 in the unit of 
$2 \pi$ which is usually adopted in the literature.  
This is known to give the anti-unitary symmetry to the
system guaranteeing the correct global quantum properties
such as level statistics.
It is easy to see that the eigenvalues of the effective Hamiltonian, 
eq. (\ref{7}) are unchanged under $\vec n_z \to -\vec n_z$, 
while the phase of eigenfunctions changes. This means that 
the physics of a particle $M$ is invariant under the reversion 
of the vector potential. In fact, for 
infinitely thin flux, the direction of $\vec B$ can 
be changed arbitrarily as long as it is off the billiard plane.
This corresponds
to the gauge transformation of the vector potential.
Explicit calculation of eq.(\ref{7}) with eq.(\ref{14}) shows that 
the vector potential acts as centrifugal barrier 
with square inverse radial dependence around each diabolical points.  
That will effectively cause the wavefunctions to avoid the flux lines.

To illustrate our arguments, we show a numerical example.
The billiard boundary is chosen to be a rectangle of size 
$[L \times 1/L]$  where  $L = 1.1557$. The mass
of the light particle is set to $m = 2 \pi$, and
the heavy particle $M = 5 m$. The strength of the coupling
is chosen to be $\bar v = 10$.
In Fig. 1, the energy eigenvalue of the system is shown.
In each of the energy levels, left side is the full
calculation, and the right side is the calculation neglecting 
the effects of flux lines.  The density profile of the 
eigenstates of the heavy particle
$\left|\left\langle \vec R\right.
\left|{\Phi_{\alpha p}}\right\rangle\right|^2$ 
for selective 
$\alpha$ and $p$ are depicted in Fig. 2,
along with the location
of the magnetic flux which is shown as black squares.
The repulsive effect of the magnetic flux line is clearly visible.

We now turn to the three dimensional case.
We consider a generic case of three-dimensional box-shaped
domain in which unperturbed single-particle spectra have 
no degeneracies.
The diabolical locations are determined by 
the same eq. (\ref{11}) as in two dimensional case, but 
with $\vec R^*$ now representing a 
three dimensional coordinate vector.
The condition, eq. (\ref{11}) specifies
a one-dimensional line
in the space of $\vec R$.  This is to be expected also from
another perspective.
Since the degeneracy is co-dimension two for real symmetric 
Hamiltonian, $\vec R^*$ cannot be 
the isolated points in the three-dimensional
space of $\vec R$.  Rather, it becomes a line of degeneracies.
Clearly, the line cannot have the edge inside the billiard
domain.    Also, the line does not go through the domain wall, since
this would mean that the unperturbed system has a degeneracy, which
we have excluded in our assumption.
The only possible configuration for the line of degeneracy,
therefore, is the {\em closed string}.
When one makes adiabatic circular motion of $\vec R$, the state 
of the sub-system $\left|{\alpha(\vec R)}\right\rangle$ obtains 
a phase $\pi$ if the circle is such that degeneracy strings 
pierce through its area odd number of times.
Otherwise, the geometric phase
is zero.  With the Stokes theorem, it is easy to show that the
gauge potential $\vec A_\alpha$ whose circular line integral
has such properties gives the flux $\vec B_\alpha$
which is non-zero only along the degeneracy string.
Thus the string of degeneracy might be called the 
{\em flux string}.
The strength of the flux $\vec B_\alpha$ along the string is 1/2 in
the unit of $2 \pi$.  Formally, it is given by
\begin{eqnarray}
\label{16}
\vec B_\alpha(\vec R)= \left( {1 \over 2} \right)\int
    {d\vec R^S\delta (f_n(\vec R^S))\delta (g_n(\vec R^S))
    \vec n(\vec R^S)2\pi \delta ^3(\vec R-\vec R^S)}
\end{eqnarray}  
where $\vec n(R)$ is the direction vector of the flux string.
As in dimension two, one can reverse the direction of 
$\vec n$ without changing the physics.

As in two dimension, the flux string in three dimension will have 
observable effects on the motion of the slow particle, and thus
on the energy spectra of the system.
Such calculation tends to become tedious for three dimension.  
Instead, in Fig. 3, we simply display the flux string 
for the system of two particles in a box of size 
$[ L_x, L_y, 1/(L_x L_y) ]$ where $L_x = 1.1557$ and $L_y = 1.0310$. 
The mass and the coupling constant are chosen to be
$m = 2 \pi$ and $\bar v = 10$ as before. 
It is evident that the flux strings
possess the pictorial nature which is quite striking. 
We think that the impact is enhanced by the fact
that these objects emerge from nothing more than 
two interacting particles in a box.
If the observation of
wave functions in a box becomes possible, these flux strings will
be visible as the empty void of the probability distribution.

The magnetic flux string similar to  the one found here
has been predicted and discussed in the context of
field theories.  In elementary particle physics, 
it is known as {\em U(1) vortex} \cite{NO73}, and in cosmology 
as {\em cosmic string} \cite{HA90}.  
It is quite conceivable that the actual experimental
observation of magnetic flux string may be first realized
at mesoscopic scale in a system similar to the one
considered here.
Further, the present study with a very
simple setting suggests that, despite its exotic appearance,
the flux string is
ubiquitous in many-body quantum systems.
We conclude that the quantum billiards acquire considerable 
richness with the introduction of more than one moving particle.
\\

We acknowledge the helpful discussions with
the members of the Theory Division of Institute
for Nuclear Study, University of Tokyo.
This work was supported in part by
the Grant-in-Aid for
Encouragement of Young Scientists (No. 07740316) 
awarded to one of us (TS)
by the Ministry of Education,
Science, Sports and Culture of Japan.
%
%

%
%
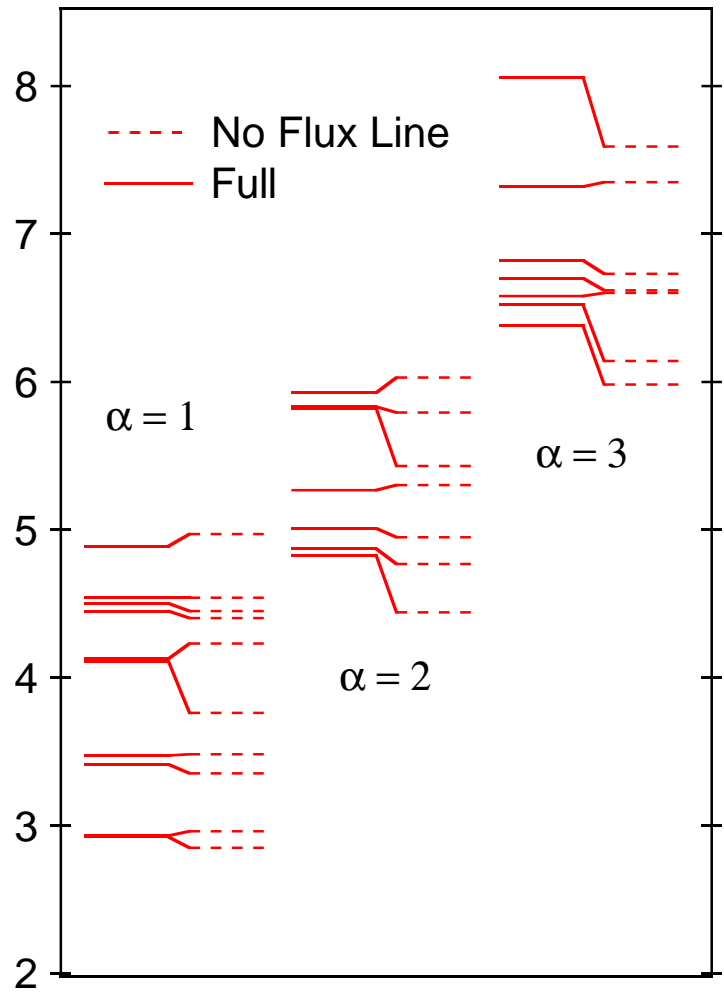
\begin{figure}
\caption{
Energy level scheme of the eigenvalue equation
eq. (10) with the effective Hamiltonian, eq. (7). 
The size of the rectangular boundary is set to
$[1.15572, 0.86526]$.  Other parameters are $m = 2\pi$,
$M = 10\pi$ and $\bar v = 10$.
}
\end{figure}
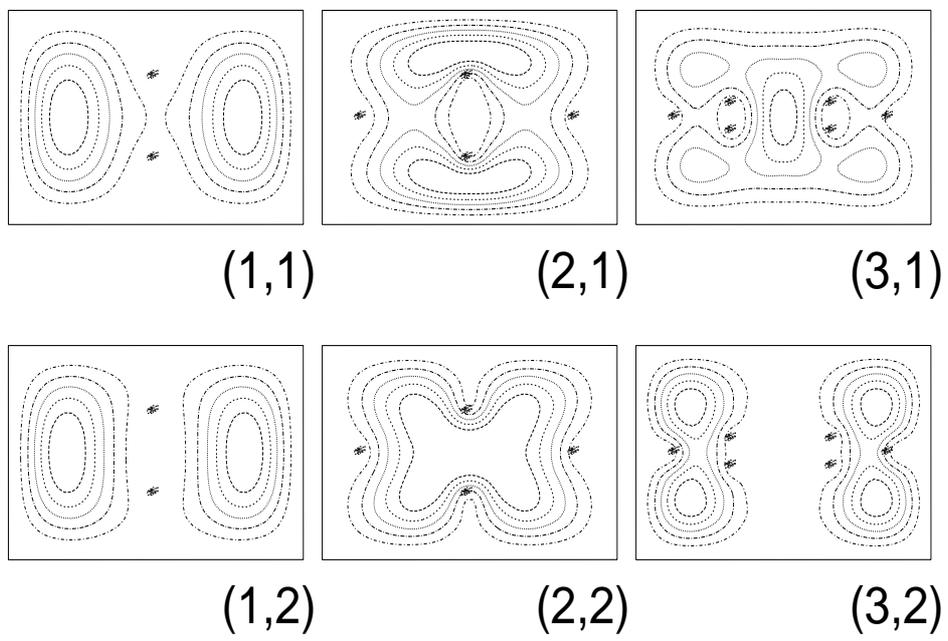
\begin{figure}
\caption{
The density profiles of the probability distribution of the
heavy particle of the several selective eigenstates of 
eq. (10) for the rectangular billiard of 
$L = 1.15572$.  the numbers in the figure refers to the 
quantum numbers $(\alpha, p)$.
}
\end{figure}
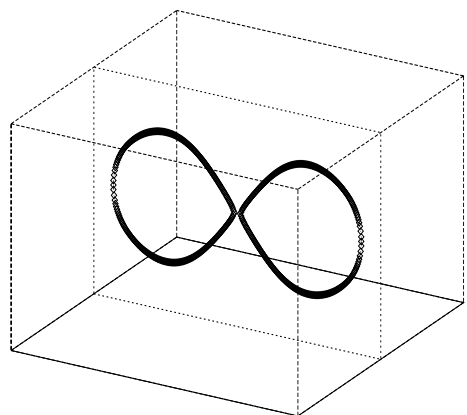
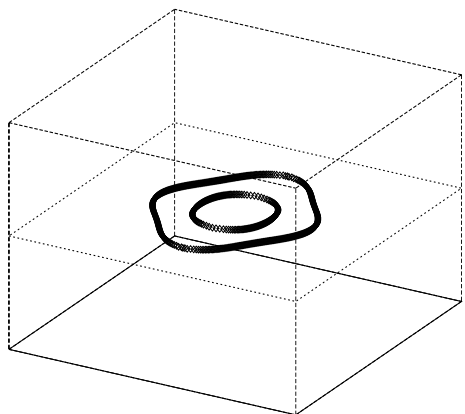
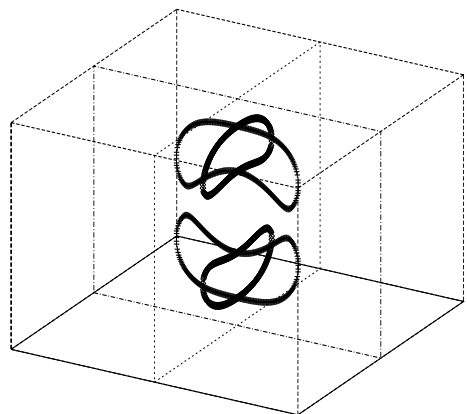
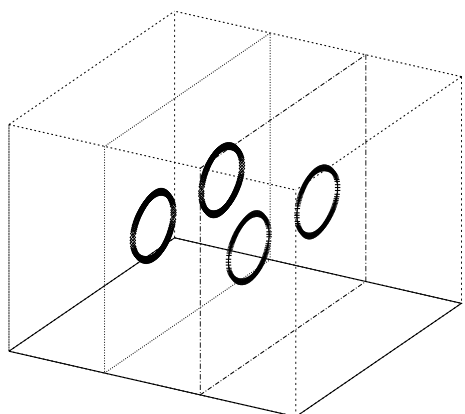
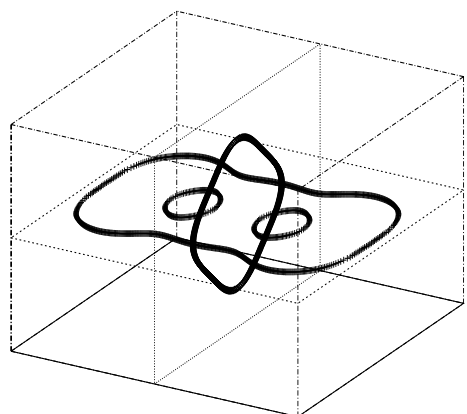
\begin{figure}
\caption{
Flux strings found in the two particle billiard problem
in a three dimensional box of the size 
$[1.1557, 1.0310, 0.8392]$.
(a) $n$ = 2, (b) $n$ = 3, (c) $n$ = 4, 
(d) $n$ = 5, (e) $n$ = 6, (f) $n$ = 7, in eq. (11), respectively. 
}
\end{figure}
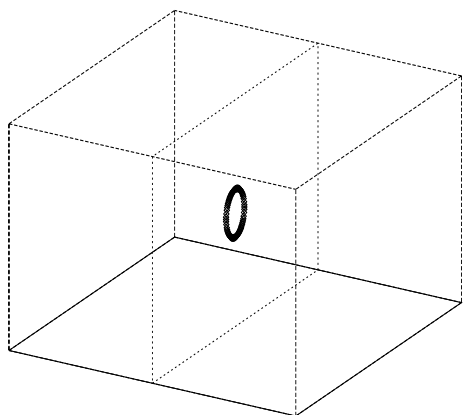
\end{document}